\documentclass[twocolumn,aps,showpacs,showkeys,prb,superscriptaddress,reprint]{revtex4-1}
\usepackage{graphicx}
\usepackage{amsmath}
\usepackage{amssymb}
\usepackage{amsfonts}
\usepackage{color}
\usepackage{bm}
\usepackage{hyperref}

\hypersetup{
	colorlinks = true,
	pdftitle = {},
	pdfauthor = {},
	pdfkeywords = {},
	linkcolor = blue,
	citecolor = blue,
	filecolor = black,
	urlcolor = magenta
}

\newcommand{\lz}{\langle \hat{L}_z \rangle}
\newcommand{\qmax}{q_\text{max}}
\newcommand{\Vacc}{V_\text{acc}}

\allowdisplaybreaks

\begin{document}

\title{Achieving atomic resolution magnetic dichroism by controlling the phase symmetry of an electron probe}

\author{J\'{a}n Rusz}
\affiliation{Department of Physics and Astronomy, Uppsala University, P.O. Box 516, 75120 Uppsala, Sweden}
\author{Juan-Carlos Idrobo}
\affiliation{Center of Nanophase Materials Sciences, Oak Ridge National Laboratory, Oak Ridge, TN 37831, USA}
\author{Somnath Bhowmick}
\affiliation{Department of Materials Science and Engineering, Indian Institute of Technology, Kanpur 208016, India}

\begin{abstract}
The calculations presented here reveal that an electron probe carrying orbital angular momentum is just a particular case of a wider class of electron beams that can be used to measure electron magnetic circular dichroism (EMCD) with atomic resolution.  It is possible to obtain an EMCD signal with atomic resolution by simply breaking the symmetry of the electron probe phase distribution using the aberration-corrected optics of an scanning transmission electron microscope.  The required phase distribution of the probe depends on the magnetic symmetry and crystal structure of the sample.  The calculations indicate that EMCD signals utilizing the phase of the electron probe are as strong as those obtained by nanodiffraction methods.
\end{abstract}


\maketitle

Development of quantitative magnetic characterization techniques goes hand-in-hand with progress in nano-technology. A terabit per square inch recording density \cite{yang,mallary} means that the area available for one bit is not larger than a square of size $25 \times 25$~nm$^2$, assuming bits arranged laterally. This pushes demands for magnetic measurements down to few nm scale \cite{bigioni,park}, approaching atomic resolution.

An attractive option to measure magnetism at such high spatial resolutions is an experimental technique based on electron magnetic circular dichroism \cite{nature,lacdif,emcd2nm,nanostuff,klie,bacteria,nanozno,nanofe3o4,cro2,fe3o4chan,polyemcd} (EMCD). Particularly, a great promise came recently from utilizing electron vortex beams \cite{bliokh,vorttem,mcmorran,vortjo,vortatom} (EVBs) within an electron microscope. With EVBs it should be possible to measure EMCD in the direction of the transmitted beam \cite{vortjo,tendeloo,lloydprl,vortexelnes,yuan,schattnp,vortexsurvey}, which brings a substantial increase in signal to noise ratio compared to intrinsic EMCD measured in between Bragg spots \cite{nature,emcd2nm,polyemcd}.  However, obtaining isolated atomic-size EVBs that can be used for EMCD measurements have not yet been possible, although different electron optical setups have been proposed \cite{jospiral,mcmorran,strongvortex,clarkaberr,saitoh2,krivanek}.

In this Letter, we show how EMCD signals can be measured with atomic resolution in the electron microscope at the transmitted beam without the necessity of producing electron probes carrying orbital angular momentum (OAM).  The calculations presented here reveal that EVBs carrying OAM are just a particular case of a wider class of electron beams that can be used to measure EMCD signals. The key feature to obtain magnetic dichroism with atomic resolution in an electron microscope is the relation between the crystal structure and magnetic symmetry of the sample, and the distribution of the phase in the electron beam. The calculations indicate that the strength of the EMCD signal is only about half of what it was reported in the first EMCD experiment on an iron crystal using a parallel beam \cite{nature}, but with the main difference that it achieves atomic spatial resolution. In consequence, electron beams that can be obtained by aberration-corrected scanning transmission electron microscopes (STEMs) without additional apertures are predicted to lead to a nonzero EMCD signal at transmitted beam.

\begin{figure}[tb]
  \includegraphics[width=8.6cm]{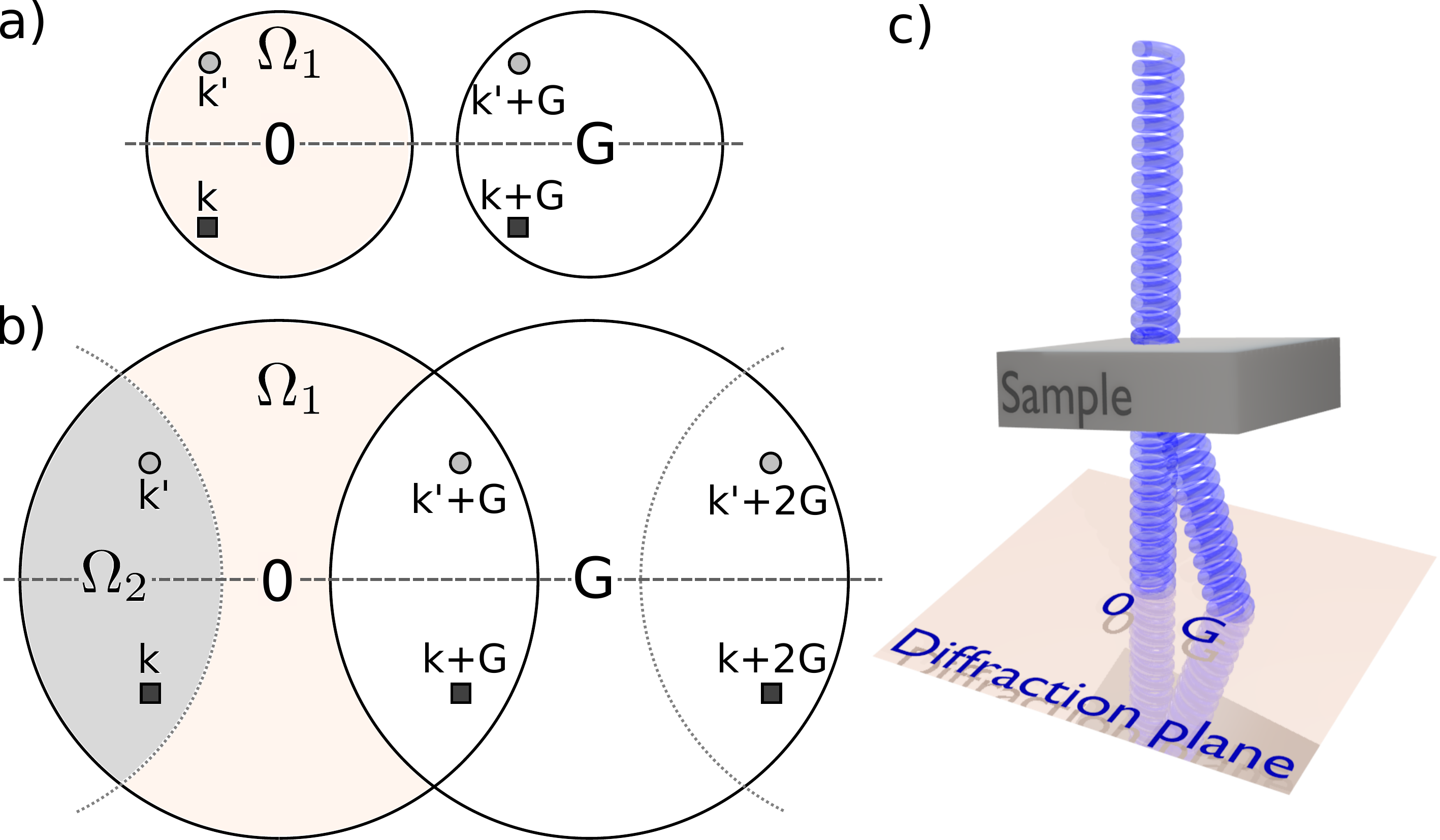}
  \caption{Schematic drawing of a diffraction of EVB assuming a) no overlap between the CBED discs, b) partial overlap of the CBED discs. Two-beam case orientation, illustrated in c), is assumed and a general $\mathbf{k}$-vector is marked, together with its mirror image $\mathbf{k'}$. The mirror axis is marked as the dashed horizontal line.}
  \label{fig:scheme}
\end{figure}

The theoretical prediction is based in a two-beam condition model for a convergent beam electron diffraction (CBED) in the STEM, as shown schematically in Fig.~\ref{fig:scheme}.  The two disks represent a transmitted beam and elastically scattered beam with Bragg vector $\mathbf{G}=(G,0,0)$.  For simplicity the model assumes only one Bragg-scattered beam, a situation with a single symmetry plane -- \emph{i.e.}, the $x$-axis. However, the results obtained here can easily be generalized to a situation with more Bragg-scattered beams and different symmetries. 

Overlap of the two CBED disks means that two regions describing the elastically scattered beam wavefunction need to be considered. Region $\Omega_1$ is such that for wavevectors $\mathbf{k} \in \Omega_1$ there is no other wavevector \emph{within the same transmitted disk}, which would differ from $\mathbf{k}$ by $\mathbf{G}$.  Contrary to $\Omega_1$, the shaded lens-shaped region $\Omega_2$ in Fig.~\ref{fig:scheme} contains the wavevectors $\mathbf{k}$, for which $\mathbf{k+G}$ lies within the same transmitted disk.  The reasons for this distinction will become obvious below.  Here however, we note that it is related to the necessary condition for atomic resolution in STEM, which requires overlap of discs to achieve a coherent interference of beam components \cite{stembook}. Under these assumptions the complete wavefunction of the elastically scattered incoming probe can be written as 
\begin{eqnarray}\label{eq:probe2}
  \psi_i(\mathbf{r}) & = & \sum_{\mathbf{k} \in \Omega_1 \cup \Omega_2 } Ce^{i\phi_\mathbf{k}}e^{i\mathbf{k}\cdot\mathbf{r}}\left[ 1 + iT_\mathbf{G}e^{i\mathbf{G}\cdot\mathbf{r}} \right] \nonumber \\
  & + & \sum_{\mathbf{k} \in \Omega_2} Ce^{i\phi_\mathbf{k+G}}e^{i(\mathbf{k+G})\cdot\mathbf{r}} \left[ 1  + i T_\mathbf{G} e^{i\mathbf{G}\cdot\mathbf{r}} \right],
\end{eqnarray}
where $C$ is a real-valued normalization constant. Bragg-scattered beam is phase shifted by $\frac{\pi}{2}$ and thus its relative amplitude can be written as $iT_\mathbf{G}$ with real-valued $T_\mathbf{G}$. Both $C$ and $T_\mathbf{G}$ are assumed to be $\mathbf{k}$-independent, which is a good approximation for thin samples usually studied in aberration-corrected STEM and spectrum imaging experiments \cite{stembook}.  The $\phi_\mathbf{k}$ represents the phase of the beam component with wavevector $\mathbf{k}$. Nonzero $\phi_\mathbf{k}$ can originate for example from aberrations or probe displacement.

For an EVB with OAM $\lz=m\hbar$ one can write $\phi_\mathbf{k}=m\arctan \frac{k_y}{k_x}$. The radius of the CBED disks is $\qmax$ which is related to the convergence semiangle $\alpha$ via $\alpha = \qmax \lambda(\Vacc)$, where $\lambda(\Vacc)$ is the de Broglie wavelength of electrons accelerated by voltage $\Vacc$.

For the outgoing wave the elastic scattering of the probe will be neglected.  Additionally, the detector will be considered to be far away, observing a single plane-wave $\psi_f(\mathbf{r}) = e^{i\mathbf{k}_f \cdot \mathbf{r}}$.

The double-differential scattering cross-section can be then evaluated as (see Supplementary Information):
\begin{eqnarray}\label{eq:ddscs}
  \lefteqn{\frac{\partial^2\sigma}{\partial \Omega \partial E} = C^2 \!\!\!\!\!\! \sum_{\mathbf{k} \in \Omega_1 \cup \Omega_2 } \!\!\!\!\!\! \Big[ S(\mathbf{q},\mathbf{q},E) + T^2_{\mathbf{G}} S(\mathbf{q-G},\mathbf{q-G},E) } \nonumber \\
  & + & 2 T_{\mathbf{G}} \mathrm{Im}[S(\mathbf{q},\mathbf{q-G},E)] \Big] \nonumber \\
  & + & \sum_{\mathbf{k} \in \Omega_2} C^2 \Big[ [1+2T_{\mathbf{G}}\sin(\Delta\phi_{\mathbf{k,G}})]S(\mathbf{q-G},\mathbf{q-G},E) \nonumber \\
  & + & T^2_{\mathbf{G}}S(\mathbf{q-2G},\mathbf{q-2G},E) \nonumber \\
  & + & 2T_\mathbf{G} \mathrm{Im}[S(\mathbf{q-G},\mathbf{q-2G},E)] \nonumber \\
  & + & 2\mathrm{Re}[e^{-i\Delta\phi_{\mathbf{k,G}}} S(\mathbf{q},\mathbf{q-G},E)] \nonumber \\
  & + & 2T_{\mathbf{G}} \mathrm{Im}[e^{-i\Delta\phi_{\mathbf{k,G}}} S(\mathbf{q},\mathbf{q-2G},E)] \nonumber \\
  & + & 2T_{\mathbf{G}}^2 \mathrm{Re}[e^{-i\Delta\phi_{\mathbf{k,G}}} S(\mathbf{q-G},\mathbf{q-2G},E)] \Big],
\end{eqnarray}
where $\Delta\phi_{\mathbf{k,G}}=\phi_{\mathbf{k+G}}-\phi_\mathbf{k}$, and
\begin{equation} \label{eq:mdff}
S(\mathbf{q},\mathbf{q'},E)=\sum_{I,F}\langle F | \frac{e^{-i\mathbf{q}\cdot\mathbf{r}}}{q^2} | I \rangle \langle I | \frac{e^{i\mathbf{q'}\cdot\mathbf{r}}}{q'^2} | F \rangle\delta(E-E_F+E_{I}),
\end{equation}
is the mixed dynamical form-factor (MDFF), with momentum transfer $\mathbf{q}=\mathbf{k}_f-\mathbf{k}$ carrying the $\mathbf{k}$-dependence of the terms in the sum in Eq.~(\ref{eq:ddscs}). The $|I\rangle,|F\rangle$ denote initial and final states of crystal of energy $E_I,E_F$, respectively. Note that the Coulomb factors have been included directly into the definition of the MDFF. The magnetic signal in the electron scattering originates from the imaginary part of MDFF according to the dipole model \cite{oursr,polyemcd}
\begin{equation}\label{eq:mdffmodel}
  S(\mathbf{q},\mathbf{q'},E) \approx \frac{N(E) \mathbf{q}\cdot\mathbf{q'} + i(\mathbf{q}\times\mathbf{q'})\cdot \mathbf{M}(E)}{q^2 q'^2},
\end{equation}
where $N(E)$ stands for isotropic non-magnetic white-line component and $\mathbf{M}(E)$ is a vector representing the magnetic component. Assuming magnetization only along $z$ direction and having $\mathbf{G}=(G,0,0)$ one obtains
\begin{equation} \label{eq:mdffmz}
  \mathrm{Im}[S(\mathbf{q}-m\mathbf{G},\mathbf{q}-n\mathbf{G},E)] = \frac{(n-m)GM_z(E)q_y}{|\mathbf{q}-m\mathbf{G}|^2|\mathbf{q}-n\mathbf{G}|^2}.
\end{equation} 


In the simplest case, when elastic scattering is neglected ($T_\mathbf{G}=0$) and the convergence angle is small, \emph{i.e.}, $\qmax < \frac{G}{2}$ and thus $\Omega_2$ is empty, Eq.~(\ref{eq:ddscs}) reduces to a simple sum of dynamical form-factors $\frac{\partial^2\sigma}{\partial \Omega \partial E} = C^2 \sum_{\mathbf{k} \in \Omega_1} S(\mathbf{q},\mathbf{q},E)$. In other words, Eq.~(\ref{eq:ddscs}) becomes an incoherent summation over all components $\mathbf{k}$ of the convergent electron probe.  No magnetic signal can arise from such condition because the dynamical form-factor is real [see Eq.~(\ref{eq:mdffmz}) for $n=m=0$].  Notice that EMCD is defined as the result of subtracting two sets of electron energy-loss spectra collected with different electron phases \cite{nature}.  If there is not a magnetic signal in the inelastic scattering, there would not be an EMCD signal either.

Assuming non-negligible elastic scattering (nonzero $T_\mathbf{G}$) for a probe with a small enough convergence angle, so that there is no overlap of the diffracted discs (empty $\Omega_2$, see Fig.~\ref{fig:scheme}a), the Eq.~(\ref{eq:ddscs}) reduces to the first sum only. The third term in the first sum explicitly contains an imaginary part of MDFF, \emph{i.e.}, one can expect a magnetic signal to be present at some scattering angles.  However, there is no dependence of the scattering cross-section on the phase $\phi_\mathbf{k}$, which means that the OAM, or in fact, any $\mathbf{k}$-space distribution of the phase in the probe does not matter.  In other words, if the convergence angle is small enough such that $\qmax < \frac{G}{2}$, the beam vorticity does not influence the inelastic scattering cross-section.

Yet, one can still observe a magnetic signal in the setting described above.  However, the distribution of the magnetic signal is antisymmetric with respect to the mirror axes. Because the mirror axes necessarily pass through the transmitted beam, an EMCD signal cannot be observed by a detector centered on the transmitted beam, regardless of how large the collection angle is.

An example of a CBED diffraction pattern calculated for an EVB with $\qmax < \frac{G}{2}$ is shown below in Fig.~\ref{fig:qmaxmaps}, left column. The proof of an antisymmetry of EMCD signal proceeds in the following way: Let's consider a wavevector $\mathbf{k'}=(k_x,-k_y,k_z)$, which is a mirror image of the wavevector $\mathbf{k}=(k_x,k_y,k_z)$, see Fig.~\ref{fig:scheme}a. Their combined contribution to the magnetic signal at $\mathbf{k}_f^{(1)}=(k_x^f,k_y^f,k_z^f)$ is evaluated using Eqn.~(\ref{eq:mdffmz}) as
\begin{equation}\label{eq:add}
  2T_\mathbf{G}GM_z(E)\left[\frac{q_y}{|\mathbf{q}|^2|\mathbf{q-G}|^2} + \frac{q'_y}{|\mathbf{q'}|^2|\mathbf{q'-G}|^2}\right],
\end{equation}
where $q_y=k_y^f-k_y$ and $q'_y=k_y^f+k_y$. Moving the detector orientation to its mirror image $\mathbf{k}^{(2)}_f = (k_x^f,-k_y^f,k_z^f)$ leads to
\begin{eqnarray}
q_y^{(2)} = -k_y^f-k_y = -q'_y \quad & \to & \quad |\mathbf{q}^{(2)}| = |\mathbf{q'}|, \label{eq:qqpr} \\
{q'}_y^{(2)}= -k_y^f+k_y = -q_y  \quad & \to & \quad |\mathbf{q'}^{(2)}| = |\mathbf{q}|, \label{eq:qprq}
\end{eqnarray}
and similarly for $|\mathbf{q-G}|$. The $q_y$ and $q'_y$ swap and change sign, \emph{i.e.}, the magnetic signal in Eqn.~(\ref{eq:add}) changes sign as well.  This holds true for all $\mathbf{k}$ from the lower half-circles of the CBED disks, thus an EMCD signal is indeed distributed \emph{antisymmetrically} with respect to the mirror axis. In particular, it vanishes right at the symmetry axis. All observations of an EMCD signal done so far, possibly except for Ref.~\onlinecite{vortjo}, are of this nature -- so called intrinsic EMCD, caused by coherence of elastically scattered beam components \cite{nature}.

\begin{figure}[htb]
 \includegraphics[width=8.6cm]{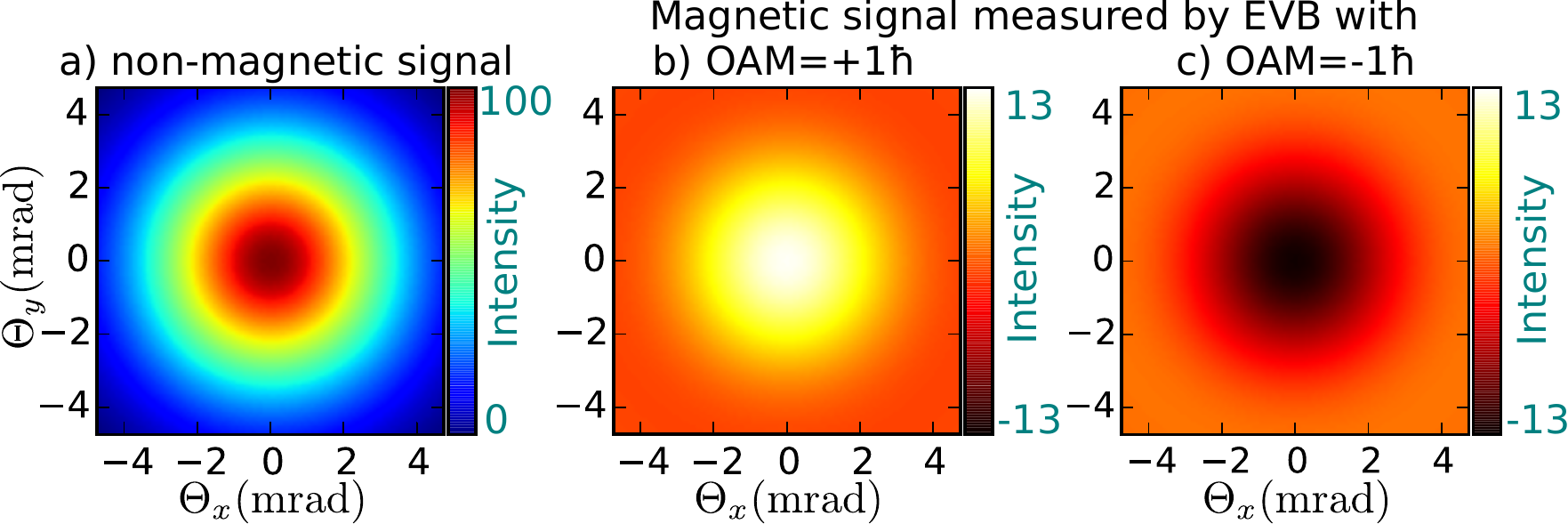}
 \caption{$L_3$-edge energy-filtered diffraction pattern of an EVB scattering on a single Fe atom. The individual panels show the non-magnetic and magnetic components (EVBs with OAM~$=\pm 1\hbar$) of the inelastic scattering cross-section assuming the Fe magnetic moment along the $z$-axis. The EMCD signal results from subtracting the OAM~$=+1\hbar$ from OAM~$=-1\hbar$ magnetic components.}
 \label{fig:singleFe}
\end{figure}

Next we consider a situation with no elastic scattering ($T_\mathbf{G}=0$), but $\qmax > \frac{G}{2}$. This can occur either for ultra-thin samples and sufficiently large convergence angles, or when the unit cell is large (\emph{i.e.}, the reciprocal lattice vectors $G$ are small).  A limiting case is a single atom in a cell with infinite lattice constant. Then for any nonzero $\alpha$ the $\qmax = \alpha / \lambda$ will be larger than $\frac{G}{2} = \frac{1}{2a} \to 0$.

The inelastic scattering cross-section can be written as
\begin{eqnarray}\label{eq:noelast}
  \frac{\partial^2\sigma}{\partial \Omega \partial E} & = & C^2 \!\!\!\!\!\! \sum_{\substack{\mathbf{k} \\ k_\perp<\qmax}} \!\!\!\!\!\! S(\mathbf{q},\mathbf{q},E) \nonumber \\
  & + & 2C^2 \sum_{\mathbf{k} \in \Omega_2} \mathrm{Re} [e^{-i\Delta\phi_{\mathbf{k,G}}} S(\mathbf{q},\mathbf{q-G},E)].
\end{eqnarray}
There are two key findings: 1) the scattering cross-section depends on the distribution of the phase in the beam wave front, 2) an EMCD signal can be observed, despite that the elastic scattering of the probe was neglected. This time, however, the imaginary part of MDFF is multiplied by a $Sine$ function of $\Delta\phi_\mathbf{k,G}$. A combined contribution of the probe component $\mathbf{k}$ and its mirror image $\mathbf{k'}$ is
\begin{equation}\label{eq:addsin}
  2T_\mathbf{G}GM_z(E)\left[\frac{\sin(\Delta\phi_\mathbf{k,G})q_y}{|\mathbf{q}|^2|\mathbf{q-G}|^2} + \frac{\sin(\Delta\phi_\mathbf{k',G})q'_y}{|\mathbf{q'}|^2|\mathbf{q'-G}|^2}\right].
\end{equation}
Note that for a vortex beam passing directly through an atomic column $\Delta\phi_\mathbf{k,G} = -\Delta\phi_\mathbf{k',G}$, which allows to take the $Sine$ function out of the brackets together with a change of the plus sign into a minus sign in between the two terms. Moving the detector from $\mathbf{k}_f^{(1)}$ to its mirror image $\mathbf{k}_f^{(2)}$ transforms the momentum transfer vectors as in Eqns.~(\ref{eq:qqpr}) and (\ref{eq:qprq}) -- both two terms change sign and then swap their order. But because of the minus sign in between them, the resulting contribution is the same at both detector orientations. This inelastic electron diffraction situation is very different from the previous case, because here there is a \emph{symmetric} distribution of EMCD with respect to the mirror axis. A symmetric distribution of the magnetic signal allows to detect EMCD at the transmitted beam. This result is in agreement with predictions of Refs.~\onlinecite{lloydprl,yuan} and recent simulations \cite{schattnp}.

The prediction can be illustrated by a simulation of the distributions of the non-magnetic and magnetic contributions to the scattering cross-section of a single Fe atom, Fig.~\ref{fig:singleFe}. The Fe atom was placed in a cell of size $5.3 \times 5.3 \times 0.53$~nm$^3$. The simulations were done using a combined multislice / Bloch-waves approach described in Ref.~\onlinecite{vortexsurvey}, with the outgoing beam described as a single plane wave and the incoming beam was an EVB with $\lz=1\hbar$ and $\qmax=0.1$~a.u.$^{-1}$, which at $\Vacc=200$~kV means $\alpha=4.7$~mrad.

\begin{figure}[th]
  \includegraphics[width=8.6cm]{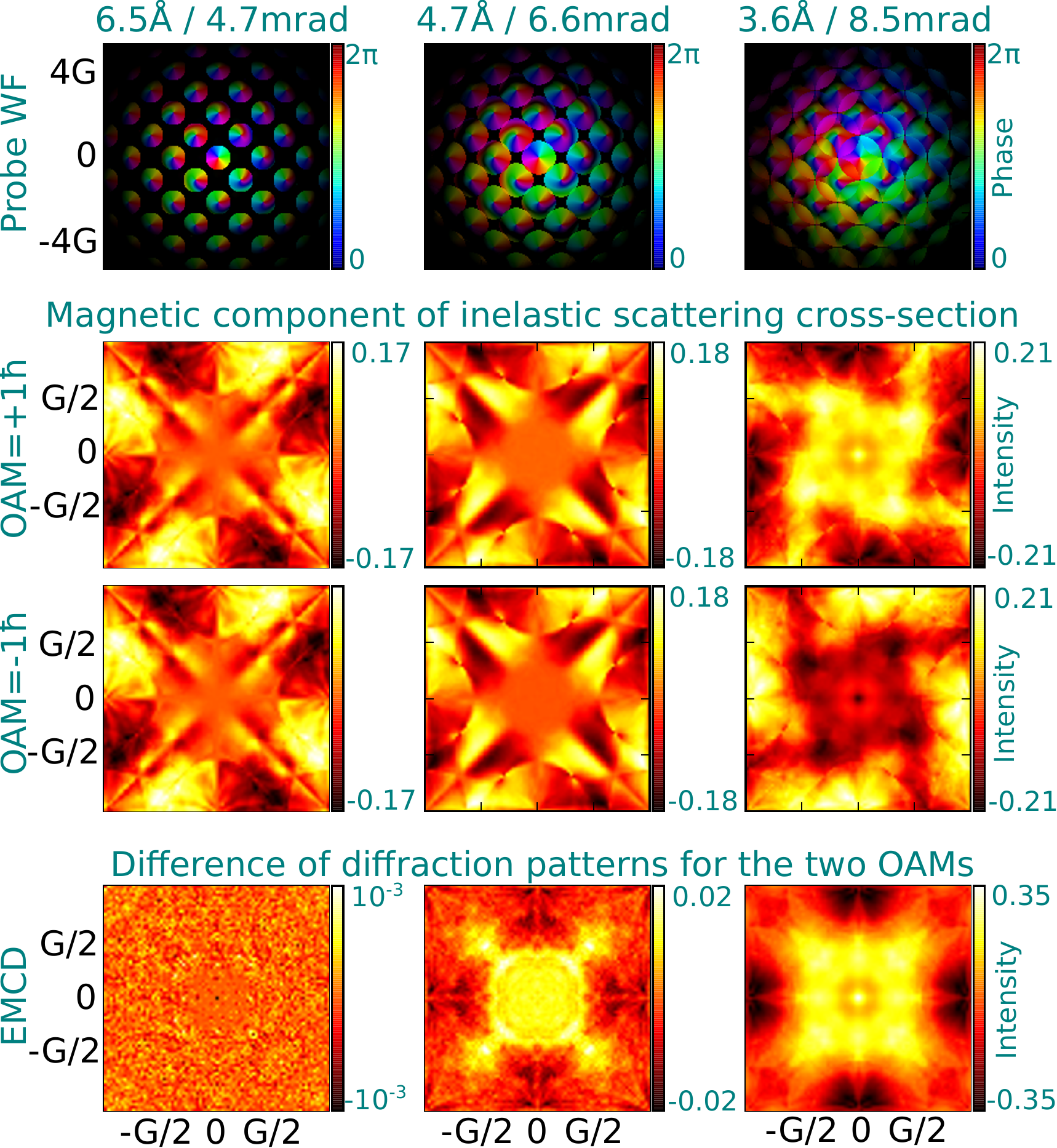}
  \caption{Dependence of probe wavefunction and energy-filtered diffraction patterns on convergence angle. (Top row) Reciprocal space probe wavefunction after passing through 10~nm of bcc iron along $(001)$ direction. Magnetic signal components of Fe-$L_3$ diffraction patterns for OAM of $\pm 1\hbar$ are shown in the middle rows. (Bottom row)  Difference of magnetic signals for the two vorticities with OAM of $\pm 1\hbar$. 
  }
  \label{fig:qmaxmaps}
\end{figure}

The calculations indicate a symmetry of the distribution of magnetic signal in diffraction plane, which is a consequence of $\Delta\phi_\mathbf{k,G} = -\Delta\phi_\mathbf{k',G}$ for all pairs of $\mathbf{k}$ and $\mathbf{k'}$ connected by a mirror symmetry.  If instead $\Delta\phi_\mathbf{k,G} = \Delta\phi_\mathbf{k',G}$, the imaginary part term in Eqn.~(\ref{eq:noelast}) would not change sign and the resulting distribution of EMCD would be antisymmetric with respect to the mirror axis.

Thus the key element here to detect an EMCD signal with atomic resolution is the $\mathbf{k}$-space distribution of the phase $\phi_\mathbf{k}$ in the electron probe.  Basically, in STEM one simply needs to set a phase distribution of the electron probe that maximizes the symmetric component of EMCD in the diffraction plane, which happens when the phase differences $\Delta\phi_\mathbf{k,G}$ are antisymmetric. Note that shifting the STEM probe from the atomic column by $\mathbf{X}$ introduces a phase factor $e^{i\mathbf{k}\cdot\mathbf{X}}$, which modifies the phase distribution and, in general, also its symmetry. In consequence, the EMCD signal intensity is reduced \cite{vortexsurvey} if the electron probe is not at the center of an atomic column; see also the Supplementary Information.

For a single atom, there is a continuum of mirror axes passing through the atom. As a consequence, the optimal beam shape for EMCD is a vortex beam passing through the atom because it has an antisymmetric phase difference with respect to all mirror axes passing through its core. 

For a crystal with a discrete set of mirror symmetries there is a wider range of phase distributions in the probe wave-front that are antisymmetric with respect to all mirror axes. This is illustrated below, where all the terms of Eq.~(\ref{eq:ddscs}) are considered.

When considering both an overlap of CBED disks ($\qmax > \frac{G}{2}$) and elastic scattering of the incoming electron probe ($T_\mathbf{G} \ne 0$), the inelastic scattering cross-section contains several terms with imaginary part of MDFF.  Some of the terms are strictly antisymmetric and do not depend on $\Delta\phi_\mathbf{k,G}$, but other terms depend on a phase difference via the real and imaginary parts of the phase factor, $e^{-i\Delta\phi_\mathbf{k,G}}$, multiplying them.  Thus an optimum phase distribution in the wave front may be rather complicated and will depend on the particular crystal structure and magnetic symmetry via dynamical diffraction effects.

With EVBs, when imaginary part of MDFF is multiplied by the real (imaginary) part of the phase factor, $e^{-i\Delta\phi_\mathbf{k,G}}$, it leads to an antisymmetric (symmetric) EMCD distribution, respectively.  It is interesting to observe how the symmetric contribution develops as the convergence angle increases.  This is illustrated in Figs.~\ref{fig:qmaxmaps} and in supplementary Fig.~1 for bcc iron crystal with $a=2.87$~\AA{}, beam direction along $(001)$ zone axis, and using an acceleration voltage of 200~kV.  The overlap onsets when $\qmax > \frac{1}{2}G_{(110)}$, where $\mathbf{G}=(110)$ is the smallest allowed reflection in a bcc structure.  Thus $\qmax$ must be larger than $\frac{1}{2}\frac{\sqrt{2}}{a}=0.1303$~a.u.$^{-1}$ so that the CBED discs will overlap and $\Omega_2$ becomes non-empty. Integrating the distribution of the magnetic signal over a circular aperture of diameter $8.7$~mrad leads to a zero magnetic signal for $\qmax \le 0.13$~a.u.$^{-1}$ and nonzero above. Clearly, when the CBED discs do not overlap (Fig.~\ref{fig:qmaxmaps}, left column), the EMCD is antisymmetrically distributed with respect to all mirror axes---horizontal, vertical and two diagonal ones---as anticipated. Above the onset of overlap this antisymmetry is broken.

\begin{figure}[htb]
 \includegraphics[width=8.6cm]{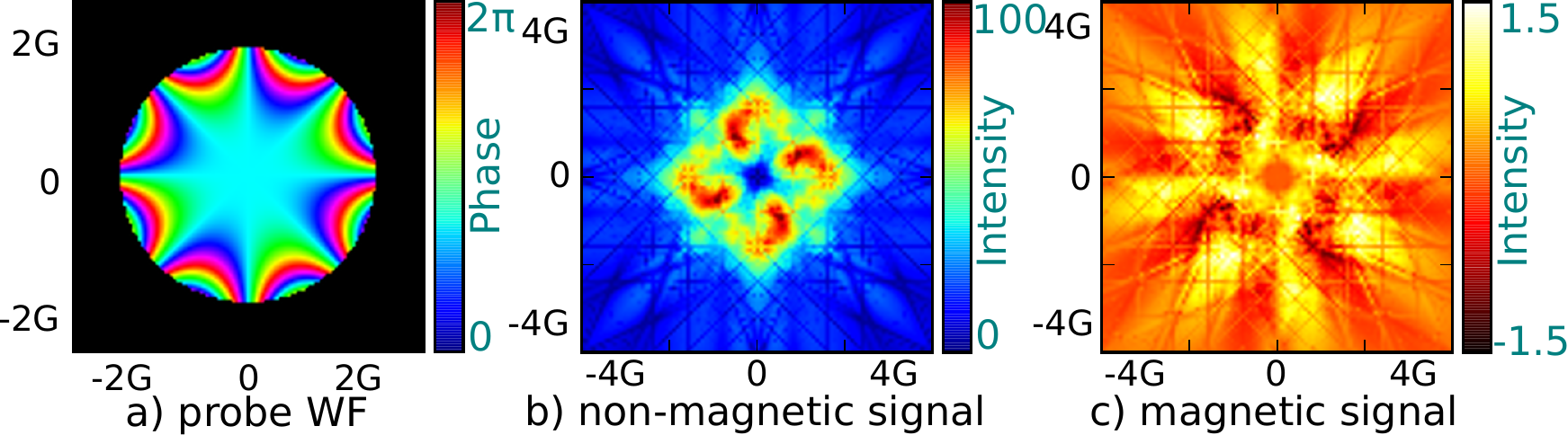}
 \caption{a) Reciprocal space wavefunction of a beam distorted by four-fold astigmatism $C_{3,4b}=0.1$~mm; $\qmax=0.4$~a.u.$^{-1}$, $\Vacc=200$~kV. b) Non-magnetic and c) magnetic component of the energy-filtered diffraction pattern at $L_3$ edge of iron, $G=1/a$.}
 \label{fig:star}
\end{figure}

Note that for a discrete set of mirror symmetries, as in the case of a bcc crystal, it is easy to construct a beam with a phase distribution, which has antisymmetric phase differences $\Delta\phi_\mathbf{k,G}$ with respect to all of the four mirror axes, but which is not a vortex beam carrying OAM.  For example, a four-fold astigmatism \cite{stembook} has the required symmetry, because for any $\mathbf{k}$ the $\phi_\mathbf{k}$ changes sign under mirror symmetry, see Fig.~\ref{fig:star}a.  An explicit calculation of the inelastic electron scattering reveals an EMCD signal distribution, which is indeed not antisymmetric.  The Fe L$_3$ signal for bcc iron integrated over a finite collection angle centered on a transmitted beam presents an EMCD of relative intensity of about 1\%. The strength of this EMCD signal is only about half of what it was reported in the first EMCD experiment \cite{nature}, but with the main difference that it has atomic spatial resolution. Moreover, with this approach one can use the full intensity of the electron beam because there is no need of a spiral \cite{jospiral,saitoh2} or fork aperture \cite{vortjo,mcmorran}, which should result in better signal to noise ratio EMCD measurements than with EVBs \cite{vortjo}.

In conclusion, we show that a finite EMCD signal with atomic resolution can be observed at the transmitted beam without the necessity of using EVBs carrying OAM.  The conditions necessary for EMCD are: 1) convergence angle large enough to cause an overlap of CBED disks and, 2) a phase distribution of the probe in reciprocal space that is not invariant under mirror symmetries of the crystal. As a corollary we propose a simple beam shape, which can be readily obtained in current aberration-corrected STEM instruments, and which should allow to measure EMCD at the transmitted beam.  The EMCD signal obtained with the method presented can achieve atomic resolution and be as strong as EMCD nanodiffraction experiments. In a more general context, our work opens up new ways of utilizing aberration-correction electron optics to design atomic-size electron probes with tailored phase distributions for specific applications and/or crystal symmetries. These electron probes could be utilized to probe magnetic dichroism, optical dichroism, and the valley polarization of materials with unmatched spatial resolution.

J.R. acknowledges Swedish Research Council and Swedish National Infrastructure for Computing (NSC center). J.-C.I. acknowledges support by the Center for Nanophase Materials Sciences (CNMS), which is sponsored at Oak Ridge National Laboratory by the Scientific User Facilities Division, Office of Basic Energy Sciences, U.S. Department of Energy.  Comments by Peter M. Oppeneer and by the referee during the review process of this manuscript are gratefully acknowledged.

\appendix

\section{Derivation of Eq.~(2)}

The double-differential scattering cross-section per scattering center (assuming $N$ atoms in the sample) can be written as
\begin{equation*}
  \frac{\partial^2\sigma}{\partial \Omega \partial E} = \frac{1}{N} \sum_{I,F} \left| \langle F, \psi_f | \hat{V} | I, \psi_i \rangle \right|^2 \delta(E-E_F+E_I),
\end{equation*}
where $|I\rangle, |F\rangle$ are the initial and final states of the crystal with energies $E_I, E_F$, and $\hat{V}$ is the Coulomb interaction potential between the nuclei and electrons of the beam and sample. The delta-function selects the transitions with energy-difference equal to energy-loss $E$.

Because of the form of the initial state of the probe wavefunction
\begin{eqnarray*}
  \psi_i(\mathbf{r}) & = & \sum_{\mathbf{k} \in \Omega_1 \cup \Omega_2 } Ce^{i\phi_\mathbf{k}}e^{i\mathbf{k}\cdot\mathbf{r}}\left[ 1 + iT_\mathbf{G}e^{i\mathbf{G}\cdot\mathbf{r}} \right] + \nonumber \\
  & + & \sum_{\mathbf{k} \in \Omega_2} Ce^{i\phi_\mathbf{k+G}}e^{i(\mathbf{k+G})\cdot\mathbf{r}} \left[ 1  + i T_\mathbf{G} e^{i\mathbf{G}\cdot\mathbf{r}} \right],
\end{eqnarray*}
the double-differential scattering cross-section contains quite a number of terms and following an explicit derivation for all of them is too lengthy and cumbersome. However, the derivation follows the same pattern for all terms. Therefore we illustrate it on a particular cross-term from the initial probe wave-function: a cross-term between $Ce^{i\phi_\mathbf{k}}e^{i\mathbf{k}\cdot\mathbf{r}}$ and $Ce^{i\phi_\mathbf{k'+G}}e^{i(\mathbf{k'+G})\cdot\mathbf{r}}$ for $\mathbf{k,k'} \in \Omega_1 \cup \Omega_2$. Essentially we will perform a similar procedure as a textbook derivation of inelastic electron scattering cross-section of a plane wave on an atom (see, e.g., Sakurai's \emph{Modern Quantum Mechanics} \cite{sakurai}) into another plane-wave $e^{i\mathbf{k}_f\cdot\mathbf{r}}$:
\begin{eqnarray*}
  \lefteqn{\frac{1}{N} \sum_{I,F} \sum_{\mathbf{k,k'}} \int d\mathbf{r} d\mathbf{x} e^{-i\mathbf{k}_f\cdot\mathbf{r}} \phi_F^\star(\mathbf{x}) \frac{1}{|\mathbf{r-x}|} \phi_I(\mathbf{x}) C e^{i\phi_\mathbf{k}} e^{i\mathbf{k}\cdot\mathbf{r}}  } \\
  & \times & \int d\mathbf{r'} d\mathbf{x'} C e^{-i\phi_\mathbf{k'+G}} e^{-i(\mathbf{k'+G})\cdot\mathbf{r'}} \phi_I^\star(\mathbf{x'}) \frac{1}{|\mathbf{r'-x'}|} \phi_F(\mathbf{x'}) e^{i\mathbf{k}_f\cdot\mathbf{r}} \\
  & \times & \delta(E-E_F+E_I) \\
  & = & \frac{1}{N} \sum_{I,F} \sum_{\mathbf{k,k'}} \int d\mathbf{r} d\mathbf{x} C e^{i\phi_\mathbf{k}} e^{-i\mathbf{q}\cdot\mathbf{r}} \phi_F^\star(\mathbf{x}) \frac{1}{|\mathbf{r-x}|} \phi_I(\mathbf{x}) \\
  & \times & \int d\mathbf{r'} d\mathbf{x'} C e^{-i\phi_\mathbf{k'+G}} e^{i(\mathbf{q'-G})\cdot\mathbf{r'}} \phi_I^\star(\mathbf{x'}) \frac{1}{|\mathbf{r'-x'}|} \phi_F(\mathbf{x'}) \\
  & \times & \delta(E-E_F+E_I)
\end{eqnarray*}
where we introduced momentum transfer vectors $\mathbf{q}=\mathbf{k}_f-\mathbf{k}$ and $\mathbf{q'}=\mathbf{k}_f-\mathbf{k'}$, which carry on the $\mathbf{k,k'}$-dependences. The crystal wavefunctions are denoted $\phi_I(\mathbf{x}),\phi_F(\mathbf{x})$ for the initial and final state, respectively. Since the $\mathbf{r,r'}$ integrations are done over the whole space, we can substitute $\mathbf{r-x}\to\mathbf{\tilde{r}}$ and analogically for the primed variables. This shift introduces phase factors $e^{-i\mathbf{q}\cdot\mathbf{x}}$ and $e^{i(\mathbf{q'-G})\cdot\mathbf{x'}}$ in the two integrals. Integration over $\mathbf{\tilde{r},\tilde{r}'}$ is then trivial and leads to
\begin{eqnarray*}
  \lefteqn{ \frac{(4\pi C)^2}{N} \sum_{I,F} \sum_{\mathbf{k,k'}} e^{i\phi_\mathbf{k}} \int d\mathbf{x} \phi_F^\star(\mathbf{x}) \frac{e^{-i\mathbf{q}\cdot\mathbf{x}}}{q^2} \phi_I(\mathbf{x}) } \\
  & \times &  e^{-i\phi_\mathbf{k'+G}} \int d\mathbf{x'} \phi_I^\star(\mathbf{x'}) \frac{e^{i(\mathbf{q'-G})\cdot\mathbf{x'}}}{|\mathbf{q'-G}|^2} \phi_F(\mathbf{x'}) \\
  & \times & \delta(E-E_F+E_I)
\end{eqnarray*}

Focusing on core-level excitations, the initial states are localized around the excited atom. Considering for simplicity crystals with only one atom per unit cell, the sum over all initial states $\sum_I$ can be written as $\sum_{\mathbf{R},\tilde{I}}$, i.e., a double sum over lattice vectors $\mathbf{R}$ and over initial states on a single atom $\tilde{I}$. Shifting the local coordinates for each $I$ by $\mathbf{R}$ we obtain
\begin{eqnarray*}
  \lefteqn{ \frac{(4\pi C)^2}{N} \sum_{\mathbf{R},\tilde{I},F} \sum_{\mathbf{k,k'}} e^{i\phi_\mathbf{k}} \int d\mathbf{x} \phi_F^\star(\mathbf{x-R}) \frac{e^{-i\mathbf{q}\cdot(\mathbf{x-R})}}{q^2} \phi_{\tilde{I}}(\mathbf{x}) } \\
  & \times & e^{-i\phi_\mathbf{k'+G}} \int d\mathbf{x'} \phi_{\tilde{I}}^\star(\mathbf{x'}) \frac{e^{i(\mathbf{q'-G})\cdot(\mathbf{x'-R})}}{|\mathbf{q'-G}|^2} \phi_F(\mathbf{x'-R}) \\
  & \times & \delta(E-E_F+E_{\tilde{I}})
\end{eqnarray*}

The final state $\phi_F(\mathbf{x})$ is an unoccupied Bloch state with wavevector $\mathbf{k}_F$ (not to be mistaken with the outgoing beam wave-vector $\mathbf{k}_f$), thus we can use a Bloch theorem stating $\phi_F(\mathbf{x-R}) = e^{i\mathbf{k}_F\cdot\mathbf{R}} \phi_F(\mathbf{x})$. That gives
\begin{eqnarray*}
  \lefteqn{ \frac{(4\pi C)^2}{N} \sum_{\mathbf{R},\tilde{I},F} \sum_{\mathbf{k,k'}} e^{i\phi_\mathbf{k}} \int d\mathbf{x} e^{-i\mathbf{k}_F\cdot\mathbf{R}} \phi_F^\star(\mathbf{x}) \frac{e^{-i\mathbf{q}\cdot(\mathbf{x-R})}}{q^2} \phi_{\tilde{I}}(\mathbf{x}) } \\
  & \times & e^{-i\phi_\mathbf{k'+G}} \int d\mathbf{x'} e^{i\mathbf{k}_F\cdot\mathbf{R}} \phi_F(\mathbf{x'}) \frac{e^{i(\mathbf{q'-G})\cdot(\mathbf{x'-R})}}{|\mathbf{q'-G}|^2} \phi_{\tilde{I}}^\star(\mathbf{x'}) \\
  & \times & \delta(E-E_F+E_{\tilde{I}})
\end{eqnarray*}
where the Bloch-wave phase factor cancels out. Introducing a notation
\begin{equation*}
  F_{F\tilde{I}}(\mathbf{q}) = \int d\mathbf{x} \phi_F^\star(\mathbf{x}) \frac{e^{-i\mathbf{q}\cdot\mathbf{x}}}{q^2} \phi_{\tilde{I}}(\mathbf{x})
\end{equation*}
the formula above can be written as
\begin{eqnarray*}
  \lefteqn{ \frac{(4\pi C)^2}{N} \sum_{\mathbf{R},\tilde{I},F} \sum_{\mathbf{k,k'}} e^{i\phi_\mathbf{k}} F_{F\tilde{I}}(\mathbf{q}) e^{i\mathbf{q}\cdot\mathbf{R}}   }\\ 
  & \times & e^{-i\phi_\mathbf{k'+G}} F_{F\tilde{I}}^\star(\mathbf{q'-G}) e^{-i(\mathbf{q'-G})\cdot\mathbf{R}} \\
  & \times & \delta(E-E_F+E_{\tilde{I}})
\end{eqnarray*}
Since $\mathbf{G}$ is a reciprocal lattice vector and $\mathbf{R}$ is a lattice vector, in the second row $e^{i\mathbf{G}\cdot\mathbf{R}}=1$. Furthermore, sum over $\mathbf{R}$ vanishes, unless $\mathbf{q-q'}$ is a reciprocal lattice vector. [More precisely, only $x$ and $y$ components of $\mathbf{q-q'}$ must be components of a reciprocal lattice vector, because the crystal has a small finite thickness. A more detailed treatment would include two Bloch waves with slightly different $z$-components of their wavevectors, leading to thickness dependent prefactors of individual MDFFs in Eq.~(\ref{eq:ddscs}) below, causing, e.g., \emph{Pendell\"{o}sung} effects. However, this does not qualitatively influence any of our conclusions.] That is equivalent to a condition requiring $\mathbf{k'}-\mathbf{k}$ to be a reciprocal lattice vector. Because of our partitioning of the CBED disks, this condition can only fulfilled when $\mathbf{k}=\mathbf{k'}$. In such case the expression simplifies to
\begin{eqnarray*}
  \lefteqn{ (4\pi C)^2 \sum_{\tilde{I},F} \sum_{\mathbf{k}} e^{i(\phi_\mathbf{k}-\phi_\mathbf{k+G})}   } \\
  & \times & F_{F\tilde{I}}(\mathbf{q}) F_{F\tilde{I}}^\star(\mathbf{q-G}) \delta(E-E_F+E_{\tilde{I}}) \\
  & \equiv & (4\pi C)^2 \sum_{\mathbf{k}} e^{-i\Delta\phi_\mathbf{k,G}} S(\mathbf{q},\mathbf{q-G},E)
\end{eqnarray*}
where $\Delta\phi_{\mathbf{k,G}}=\phi_{\mathbf{k+G}}-\phi_\mathbf{k}$ and we have introduced the mixed dynamical form-factor (MDFF)
\begin{eqnarray*}
S(\mathbf{q},\mathbf{q'},E) & = & \sum_{\tilde{I},F}\langle F | \frac{e^{-i\mathbf{q}\cdot\mathbf{r}}}{q^2} | \tilde{I} \rangle \langle \tilde{I} | \frac{e^{i\mathbf{q'}\cdot\mathbf{r}}}{q'^2} | F \rangle\delta(E-E_F+E_{\tilde{I}}) \\
 & \equiv & \sum_{\tilde{I},F} F_{F\tilde{I}}(\mathbf{q}) F_{F\tilde{I}}^\star(\mathbf{q'}) \delta(E-E_F+E_{\tilde{I}})
\end{eqnarray*}

Taking a cross-term with swapped order of the terms from initial probe wave-function would lead to
\begin{equation*}
  (4\pi C)^2 \sum_{\mathbf{k}} e^{i\Delta\phi_\mathbf{k,G}} S(\mathbf{q-G},\mathbf{q},E)
\end{equation*}
Utilizing the Hermitian property of MDFF $S(\mathbf{q},\mathbf{q'},E) = S^\star(\mathbf{q'},\mathbf{q},E)$, we obtain for their sum
\begin{eqnarray*}
  \lefteqn{ (4\pi C)^2 \sum_{\mathbf{k}} \Big[ e^{-i\Delta\phi_\mathbf{k,G}} S(\mathbf{q},\mathbf{q-G},E) } \\
  & + & e^{i\Delta\phi_\mathbf{k,G}} S(\mathbf{q-G},\mathbf{q},E) \Big] \\
  & = & (4\pi C)^2 \sum_{\mathbf{k}} 2\mathrm{Re} [ e^{-i\Delta\phi_\mathbf{k,G}} S(\mathbf{q},\mathbf{q-G},E) ] \\
  & \equiv & (4\pi C)^2 \sum_{\mathbf{k}} 2\Big[ \cos\Delta\phi_\mathbf{k,G} \mathrm{Re}[S(\mathbf{q},\mathbf{q-G},E)] \\
  & + & \sin\Delta\phi_\mathbf{k,G} \mathrm{Im}[S(\mathbf{q},\mathbf{q-G},E)] \Big]
\end{eqnarray*}
which is one of the terms in Eq.~(2) of the main text.

Chosen partitioning of the CBED disc means that nonzero cross-terms only happen, when $\mathbf{k,k'}$ are both from the same region---either $\Omega_1$ or $\Omega_2$, respectively. The final result considering all cross-terms [and dropping the constant factor $(4\pi)^2$] is then
\begin{eqnarray}
  \lefteqn{\frac{\partial^2\sigma}{\partial \Omega \partial E} = C^2 \!\!\!\!\!\! \sum_{\mathbf{k} \in \Omega_1 \cup \Omega_2 } \!\!\!\!\!\! \Big[ S(\mathbf{q},\mathbf{q},E) + T^2_{\mathbf{G}} S(\mathbf{q-G},\mathbf{q-G},E) } \nonumber \\
  & + & 2 T_{\mathbf{G}} \mathrm{Im}[S(\mathbf{q},\mathbf{q-G},E)] \Big] \nonumber \\
  & + & \sum_{\mathbf{k} \in \Omega_2} C^2 \Big[ [1+2T_{\mathbf{G}}\sin(\Delta\phi_{\mathbf{k,G}})]S(\mathbf{q-G},\mathbf{q-G}) \nonumber \\
  & + & T^2_{\mathbf{G}}S(\mathbf{q-2G},\mathbf{q-2G}) \nonumber \\
  & + & 2\cos(\Delta\phi_{\mathbf{k,G}}) \mathrm{Re}[S(\mathbf{q},\mathbf{q-G})] \nonumber \\
  & + & 2\sin(\Delta\phi_{\mathbf{k,G}}) \mathrm{Im}[S(\mathbf{q},\mathbf{q-G})] \nonumber \\
  & - & 2T_{\mathbf{G}} \sin(\Delta\phi_{\mathbf{k,G}}) \mathrm{Re}[S(\mathbf{q},\mathbf{q-2G})] \nonumber \\
  & + & 2T_{\mathbf{G}} \cos(\Delta\phi_{\mathbf{k,G}}) \mathrm{Im}[S(\mathbf{q},\mathbf{q-2G})] \nonumber \\
  & + & 2T_{\mathbf{G}}^2 \cos(\Delta\phi_{\mathbf{k,G}}) \mathrm{Re}[S(\mathbf{q-G},\mathbf{q-2G})] \nonumber \\
  & + & 2[T_{\mathbf{G}}^2 \sin(\Delta\phi_{\mathbf{k,G}}) + T_\mathbf{G}] \mathrm{Im}[S(\mathbf{q-G},\mathbf{q-2G})] \Big], 
\end{eqnarray}
equal to Eq.~(2) in the main text. A notable feature of this expression is a lack of cross-terms between different $\mathbf{k}$-vectors within a CBED disc. This is, as was shown above, due to the sum over lattice vectors $\frac{1}{N}\sum_\mathbf{R} e^{i(\mathbf{q-q'})\cdot\mathbf{R}}$, which appears due to shifts of local coordinate systems for each $I$.

\section{Shifts of the STEM probe}

It is important to note that the phase distribution in the reciprocal space, $\phi_\mathbf{k}$, can consist of several contributions. The aberrations of electron optics, such as defocus, spherical aberrations, comas or star aberrations, are among the most common contributions to the non-trivial phase distribution. They can be to a large extent removed by probe aberration correctors. Alternatively, as is suggested here, probe correctors can be utilized to deliberately set-up a desired phase distribution in the beam.

There is, however, another source of the phase variation, which can not be principially removed in a STEM experiment -- the phase ramp due to shift of the probe. A probe shifted from origin (e.g., from an investigated atomic column) by position vector $\mathbf{X}$ will introduce a phase ramp $e^{i\mathbf{k}\cdot\mathbf{X}}$ in the reciprocal space probe wavefunction, as a consequence of the Fourier shift theorem \cite{stembook}. Therefore, if the STEM probe does not pass directly through an investigated atomic column, the phase distribution will be non-trivially modified, including modification of the symmetry of the phase-distribution. This will, for example, violate the relation $\Delta\phi_\mathbf{k,G} = -\Delta\phi_\mathbf{k',G}$ valid for a vortex beam passing through an atomic column -- as discussed in relation to Eq.~(10) of the main text. It is not easy to qualitatively estimate the consequence of shift on the strength of EMCD, but explicit calculations have been performed for vortex beams with OAM=1 in Ref.~\cite{vortexsurvey}. Those can be qualitatively summarized in the following way: 1) for small shifts EMCD monotonously decreases with increasing shift $|\mathbf{X}|$ from an atomic column; 2) for $|\mathbf{X}| > \frac{1}{2}\times$FWHM of a beam with the same convergence angle, but with $m=0$, the EMCD will be reduced to less than 1/2 of the intensity for a beam passing directly through an atomic column.

\section{Simulation of EMCD strength as a function of convergence angle}

\begin{figure}[thb]
  \includegraphics[width=8.6cm]{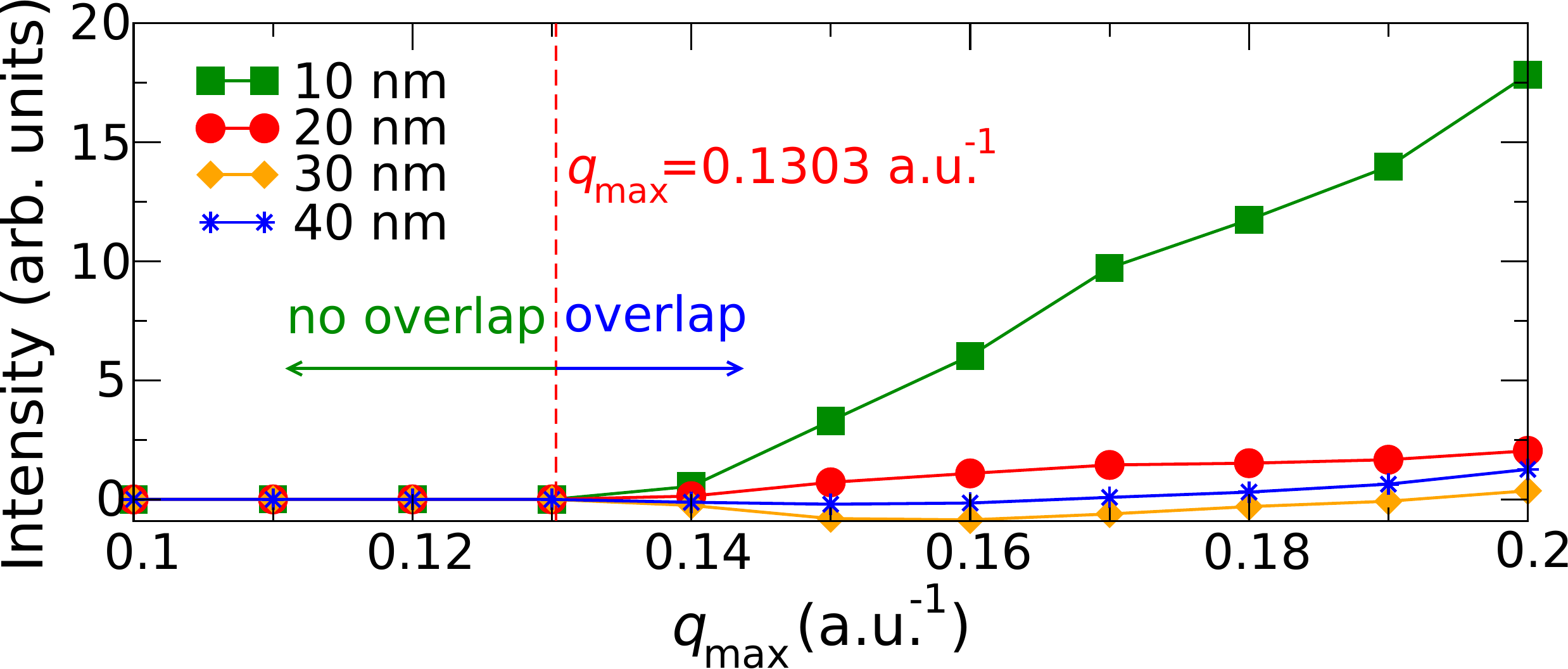}
  \caption{EMCD signal at $L_3$ edge of iron integrated over an on-axis aperture of diameter 8.7~mrad as a function of $\qmax$ and sample thickness from 10 to 40~nm. Overlap of CBED discs onsets at $\qmax=\frac{\sqrt{2}}{2a}=0.1303$~a.u.$^{-1}$.}
  \label{fig:qmaxdep}
\end{figure}

In the main text, the Fig.~3 illustrated the dependence of the EMCD signal detectable by electron vortex beams of three different diameters. In the first column, the CBED disks did not overlap ($\qmax=0.1$~a.u.$^{-1}$), in the second column there was a slight overlap ($\qmax=0.14$~a.u.$^{-1}$) giving rise to a small but nonzero EMCD at transmitted beam, and in the third column the CBED disks ovelap strongly ($\qmax=0.18$~a.u.$^{-1}$) leading to a substantial EMCD at transmitted beam direction.

Here we extend the analysis by showing the dependence of the EMCD signal as a function of CBED disk diameter $\qmax$ in the range from $0.1$ to $0.2$~a.u.$^{-1}$, see Fig.~\ref{fig:qmaxdep}. As mentioned in the main text, the overlap onsets when $\qmax > \frac{1}{2}G_{(110)}$, where $\mathbf{G}=(110)$ is the smallest allowed reflection in a bcc structure.  Thus $\qmax$ must be larger than $\frac{1}{2}\frac{\sqrt{2}}{a}=0.1303$~a.u.$^{-1}$ so that the CBED discs overlap and $\Omega_2$ is not empty. Integrating the distribution of the magnetic signal over a circular aperture of diameter $8.7$~mrad we indeed observe a zero magnetic signal for $\qmax \le 0.13$~a.u.$^{-1}$ and nonzero above, Fig.~\ref{fig:qmaxdep}.

Note that the sign of EMCD signal is not fixed by OAM of the beam and direction of the magnetic moment. Particularly, for sample thickness of 30~nm the EMCD signal is negative for a rather wide range of $\qmax$ values. This is not surprising, when one considers the origin of the vortex induced EMCD, as discussed in the main text. Dynamical diffraction effects, resulting in Pendell\"{o}sung oscillations, determine the relative size and \emph{sign} of the factors $T_\mathbf{G}$. Sign of this factor depends on the thickness of the sample and on extinction distance of the $\mathbf{G}$ beam. Thus one can not \emph{a priori} interpret the sign of observed EMCD in terms of direction of the magnetic moment and OAM of the beam alone.

\end{document}